\documentclass{Interspeech2024}
\usepackage{epsfig,amssymb,amsmath}
\usepackage{cite}
\usepackage{booktabs}
\usepackage{float}
\usepackage{multirow}
\usepackage{tipa}
\usepackage{svg}
\usepackage{adjustbox}
\DeclareMathOperator*{\argmin}{arg\,min}
\ninept

\newcommand{\phoneme}[1]{/\textipa{#1}/}
\interspeechcameraready

\title{Explainable by-design Audio Segmentation through Non-Negative Matrix Factorization and Probing\thanks{$^*$ Equal contribution.}}

\name[affiliation={1,2}]{Martin}{Lebourdais$^*$}
\name[affiliation={1,3}]{Théo}{Mariotte$^*$}
\name[affiliation={4}]{Antonio}{Almudévar}
\name[affiliation={1}]{Marie}{Tahon}
\name[affiliation={4}]{Alfonso}{Ortega}

\address{
  $^1$LIUM, Le Mans University, Le Mans, France\\
  $^2$IRIT, Toulouse, France \\
  $^3$IDS, Telecom Paris, Palaiseau, France
  $^4$ViVoLab, University of Zaragoza, Spain}
\email{marie.tahon@univ-lemans.fr}

\keywords{Audio segmentation, NMF, explainability, probing}

\begin{document}

\maketitle

\begin{abstract}
Audio segmentation is a key task for many speech technologies, most of which are based on neural networks, usually considered as black boxes, with high-level performances.
However, in many domains, among which health or forensics, there is not only a need for good performance but also for explanations about the output decision.
Explanations derived directly from latent representations need to satisfy ``good'' properties such as informativeness, compactness, or modularity, to be interpretable.
In this article, we propose an explainable-by-design audio segmentation model based on non-negative matrix factorization (NMF) which is a good candidate for the design of interpretable representations.
This paper shows that our model reaches good segmentation performances, and presents deep analyses of the latent representation extracted from the non-negative matrix.
The proposed approach opens new perspectives toward the evaluation of interpretable representations according to ``good'' properties.

\end{abstract}

\vspace{-10pt}
\section{Introduction}


Audio segmentation is a key task for many speech technologies such as automatic speech recognition, speaker identification, and dialog monitoring in different multi-speaker scenarios, including TV/radio, meetings, and medical conversations. 
More precisely, these technologies must be aware of the presence of noisy environments (brouhaha, external noise), and how many speakers are active at each time.
Indeed, the current trend for explainable AI is a vital process for transparency of decision-making with machine learning: the user (a doctor, a judge, or a human scientist) has to justify the choice made based on the system output.
Among the strategies towards explainable AI, one can find explainable-by-design models in which the explanation is not an add-on but is embedded as an essential component of the system. 
Our work comes within the scope of such models.
The explanation can be directly derived from a representation subspace of the latent representation, but ``good'' representations need to satisfy some properties \cite{bengio2013good}. Among all possible properties, modularity, compactness, and informativeness are the most important \cite{carbonneau2022metrics}.
A representation is modular if a single factor affects the subspace.
It is compact if the subspace affected by one factor is as small as possible, ideally a single dimension.
It is informative if factors can be predicted from the latent representation.

Non-negative matrix factorization (NMF) is a technique that has been widely used in audio processing \cite{bisot_overlapping_2017}, and more particularly for explainability \cite{parekh2023tackling,mariotte2024explainable}.
%
In \cite{mariotte2024explainable}, activated components extracted from the NMF trained as a post-hoc student model, discriminate sound classes.
The frequency bins used for the decision can be easily identified at both the segment (local explanations) and global levels (class prototypes).

In the present study, we propose an explainable-by-design multilabel audio segmentation based on NMF which predicts simultaneously the presence of sound classes.
We demonstrate that this model can reach good performances in Speech Activity Detection (SAD), Overlap Speech Detection (OSD), Music Detection (MD), and Noise Detection (ND).
In this article, we also provide a formal approach to analyze and evaluate three properties of the NMF components: informativeness, compactness, and modularity.
Informativeness is evaluated by probing components with classification tasks (phonemes, sound events, gender, and music styles).
Compactness and modularity are investigated with the analysis of the component structure. 
The code to reproduce the models is available at \url{https://github.com/Lebourdais/3MAS}.




\vspace{-5pt}
\section{Related works}\label{sec:related}

When processing audio data, multiple challenges arise, one of them being the diversity of information present in the audio signal. 
In the literature, segmentation tasks are often completed by separate bi-directional recurrent or convolutional models~\cite{vesperini2016vad,kim2022vad}, or Temporal Convolutional Network (TCN)~\cite{cornell_detecting_2020,lebourdais22_interspeech,bai_empirical_2018} trained on different datasets, thus increasing computational costs and limiting the usage to specific datasets.
Additionally, 
an OSD convolutional model~\cite{jung21_interspeech} deals with this problem from a multiclass perspective, while a modified version of the end-to-end diarization (EEND) approach~\cite{bredin21_interspeech} is based on the multilabel paradigm. 

Different approaches have been investigated to train models where latent representations satisfy some constraints regarding interpretability.
One of the main difficulties is to link the desired properties with mathematical constraints.
Sparse representations are commonly investigated for explainable AI since it reduces the number of dimensions to consider \cite{minh2022explainable}.
In SPINE \cite{subramanian2018spine} and Sparse-NMF (SNMF) \cite{le2015sparse}, sparsity is guaranteed by the use of a $L_1$ loss which forces values to be close to 0.
Another ``good'' property for interpretable representations is that all possible sources of variations should be disentangled \cite{bengio2013good}. 
Knowing the high diversity of audio events, we understand that this term usually covers various techniques.
The orthogonality of generative factors extracted from data is closely related to disentanglement \cite{piaggesi2023dine}.
In variational auto-encoders, the ELBO loss aims at minimizing the deviation from prior, thus imposing independence in latent dimensions \cite{higgins2017betavae}.

Information factorization aims to explicitly separate latent representation into multiple factors. 
For instance, in \cite{qian_autovc_2019} authors disentangle rhythm, pitch, timbre, and linguistic content through a factorization process within a neural network.
Recently, NMF has been extensively used for audio processing as a powerful factorization technique for structure discovery \cite{nieto_systematic_nodate}. 
NMF allows for positive and sparse \cite{le2015sparse} representations which makes them good candidates for interpretability.
In \cite{mariotte2024explainable}, NMF has been used as a student proxy model to explain the decision provided by a multilabel segmentation teacher.
The non-negative matrix has been shown to provide both global explanations in terms of relevant frequencies,
and local explanations (how much each component contributes to the final decision of a given input).


\section{Multilabel NMF segmentation model}

This section describes the NMF segmentation architecture.
It simultaneously predicts the presence of four different sound classes at the frame level. 

\begin{figure}[t]
    \centering
    \includegraphics[width=\linewidth]{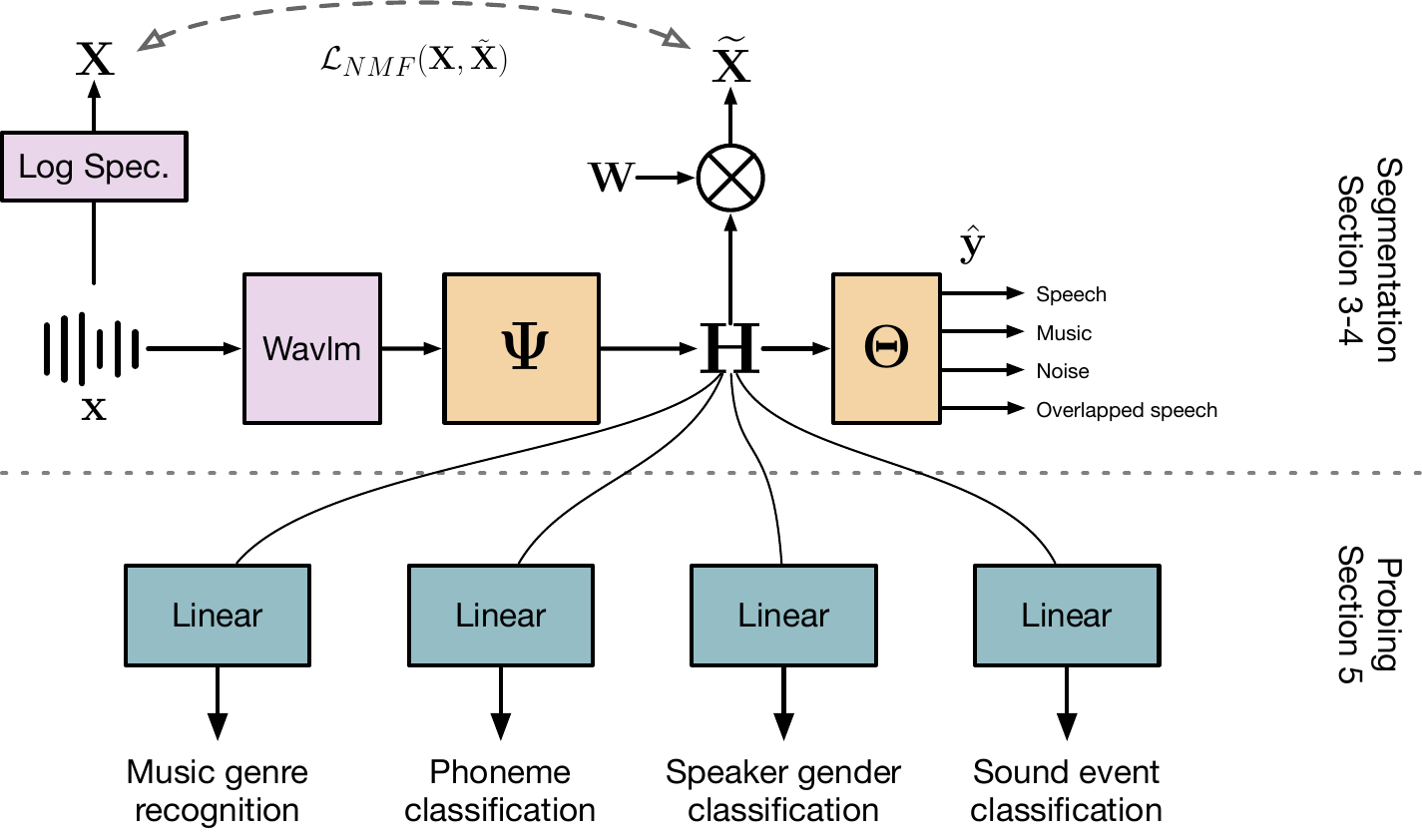}
    \caption{The 3MAS-NMF explainable by-design segmentation model (top)
    with the different probes (bottom) used to explore the informativeness of the $\mathbf{H}$ embedding. Log spec. means log-spectrogram.}
    \vspace{-15pt}
    \label{fig:archi}
\end{figure}

\subsection{Problem formulation}
The proposed architecture, presented in Fig.~\ref{fig:archi}, is inspired by \cite{parekh2023tackling} and follows the student segmentation system of \cite{mariotte2024explainable} in which the teacher is a black box model \cite{lebourdais2024multilabel}.
In this study, we remove the teacher-student training procedure to build a new explainable-by-design system.

%
%
Let $\{\mathbf{S}, \mathbf{y}\}$ be a training set composed of acoustic features $\mathbf{S}\in\mathbb{R}^{D\times T}$ extracted from an audio signal, where $D$ is the feature vector dimension and $T$ the number of frames, and the aligned annotations $\mathbf{y}\in \mathbb{R}^{C\times T}$, with $C$ the number of classes. 
For a given class $c$, the binary reference at frame $t$ verifies $y_{c,t}\in\{0,1\}$.
The feature sequence $\mathbf{S}$ is then processed by $\Psi: \mathbb{R}^{D\times T}\rightarrow \mathbb{R}_+^{K\times T}$ where $K$ is the number of factorized components. 
The $\Psi$ function extracts the non-negative embedding $\mathbf{H}\in\mathbb{R}_+^{K\times T}$.
Finally, the logits are obtained by the $\Theta$ function such as $\hat{\mathbf{y}}=\Theta(\mathbf{H})$.
%
%
The $\mathbf{H}$ embedding is designed to reconstruct the spectrogram of the input signal $\tilde{\mathbf{X}}\in\mathbb{R}_+^{F\times T}$.
This is done by multiplying $\mathbf{H}$ with a pre-trained NMF dictionary $\mathbf{W}\in\mathbb{R}_+^{F\times K}$ : $\tilde{\mathbf{X}} = \mathbf{W}\mathbf{H}$.
The $\mathbf{W}$ dictionary can be seen as a codebook of frequencies that can be activated by the embedding matrix $\mathbf{H}$ to reconstruct the spectrogram.






\subsection{Training procedure}

The segmentation model is trained with three training objectives.
The first objective is the frame classification to assign a given feature frame to a set of classes.
Since the segmentation is solved as a multilabel classification task, we use the binary cross-entropy $\mathcal{L}_{BCE}(\hat{\mathbf{y}},\mathbf{y})$.
%
The second loss constrains $\mathbf{H}$ to reconstruct the spectrogram of the input segment $\mathbf{X}$:
\begin{equation}\label{eq:nmf_loss}
    \mathcal{L}_{NMF}(\mathbf{X},\tilde{\mathbf{X}})=\lVert\mathbf{X}-\mathbf{W}\mathbf{H}\rVert_2^2
\end{equation}
Thus, the activation matrix solves both segmentation and spectrogram reconstruction.
The last loss term enforces the $\mathbf{H}$ activation matrix to be sparser with a $L_1$ norm as discussed in section \ref{sec:related}.
In the end, the global loss function can be written as:
\begin{equation}
    \mathcal{L} = \alpha \mathcal{L}_{BCE}(\hat{\mathbf{y}},\mathbf{y}) + \beta \mathcal{L}_{NMF}(\mathbf{X},\tilde{\mathbf{X}}) + \gamma \lVert\mathbf{H}\rVert_1
    \label{eq:opt_obj}
\end{equation}

\begin{table}[t!]
    \centering
    \setlength{\tabcolsep}{ 3pt } 
    \caption{Datasets used through this study with the label available \textit{SAD: speech, OSD: overlap, MD: music, ND: noise}. AragonRadio is a subset of Albayz\'in.}
    \begin{adjustbox}{max width=0.92\linewidth}
    \begin{tabular}{lccccc}
    \toprule
     Dataset & Hours & SAD & OSD & MD & ND \\
     & \textit{train/dev/test} & & & & \\
     \midrule
     DiHard III \cite{ryant2021dihard} & 25.44/8.44/32.96 & \checkmark & \checkmark & & \\
     Albayz\'in \cite{albayzin10,albayzin12} & 62.74/30.06/18.00 &\checkmark &  & \checkmark & \checkmark\\
     {\scriptsize \hspace{0.5cm} Aragon Radio} & {\scriptsize 5.28/-/18.00}& & & & \\
     ALLIES \cite{tahon2024allies} & 183.97/12.08/183.8 & \checkmark & \checkmark & & \\
     \bottomrule
    \end{tabular}
    \end{adjustbox}
    \label{tab:datasets}
    \vspace{-10pt}
\end{table}

\begin{table*}[t]
    \centering
    \caption{F1-score (\%) on each segmentation task with AragonRadio (evaluation set of Albayz\'in) and DiHard (eval full) datasets. Bold indicates the best-performing system. The confidence interval at 95\% is calculated.
    }
    \begin{tabular}{lcccccccc}
        \toprule
         & & & & \multicolumn{3}{c}{AragonRadio} & \multicolumn{2}{c}{DiHard} \\
         \cmidrule(lr){5-7}
         \cmidrule(lr){8-9}
         System & $\alpha$ & $\beta$ & $\gamma$ &  SAD & ND & MD & SAD & OSD\\  
         \midrule
          Teacher 3MAS \cite{lebourdais2024multilabel} & 1 & - & - & 96.8$\pm$0.25 & 78.6$\pm$0.58 & 93.2$\pm$0.36 & \textbf{96.9}$\pm$0.10 & 60.7$\pm$0.28\\
          Student NMF\cite{mariotte2024explainable}  & 10 & 1 & 0.1 & 96.8$\pm$0.25  & 79.5$\pm$0.57 & 93.1$\pm$0.36 & \textbf{96.9}$\pm$0.10 & \textbf{61.4}$\pm$0.28 \\
          \midrule
          Our Seg-NMF & 10 & 5 & 0.1 & 97.2$\pm$0.23 & 78.4$\pm$ 0.59 & 92.4$\pm$0.38 & 96.5$\pm$0.10 & 49.6$\pm$0.28\\
          Our Seg-NMF & 10 & 1 & 0.1 & 97.4$\pm$0.23 & \textbf{80.7}$\pm$0.56 &\textbf{93.8}$\pm$0.34 & 96.7$\pm$0.10 & 57.0$\pm$0.28\\
          Our Seg-NMF & 10 & 0 & 0.1 &\textbf{97.5}$\pm$0.22 & 79.6$\pm$0.57 & 93.3$\pm$0.36 & \textbf{96.8}$\pm$0.10 & 61.1$\pm$0.28\\
         \bottomrule
    \end{tabular}
    \label{tab:seg_pef}
    \vspace{-10pt}
\end{table*}

\vspace{-5pt}
\subsection{Implementation details}

\textbf{Acoustic features}
We use WavLM-\textit{base} \cite{chen2022wavlm} as a feature extractor (weights are frozen) to obtain the $\mathbf{S}$ feature sequence from the input audio.
This results in a sequence of 1024-dimension vectors with a 20ms framerate.
Training segments have a 4-second duration.\smallskip

\noindent \textbf{The $\Psi$ function} extracts the non-negative representation $\mathbf{H} = \Psi(\mathbf{S}) \in \mathbb{R}_+^{K\times T}$ with $K=256$.
It is composed of a 64-channel bottleneck layer followed by 3 TCN blocks \cite{bai_empirical_2018} composed of 5 1-D convolutional layers with exponentially increasing dilatation. 
A skip connection is added between TCN blocks.
A final 1-D convolution layer followed by a ReLU activation outputs the non-negative $\mathbf{H}$ embedding.\smallskip


\noindent \textbf{The $\Theta$ function} maps the $\mathbf{H}$ activation matrix to the decision space. 
Similarly to \cite{parekh2023tackling,mariotte2024explainable}, $\Theta$ is designed as a single linear layer with no bias such as $\hat{\mathbf{y}} = \boldsymbol{\theta}\mathbf{H}$,
where $\boldsymbol{\theta}\in\mathbb{R}^{C\times K}$ are the trainable weights of the linear layer. 
In our experiments, we target $C=4$ different classes.\smallskip


\noindent \textbf{The $\mathbf{W}$ dictionary} is pre-trained on a subset of the training data used for the segmentation model, with SNMF~\cite{le2015sparse}.
This consideration limits the number of activated frequencies in $\mathbf{W}$ and consequently in the activations $\mathbf{H}$.
The SNMF solves the optimization problem in eq.~\ref{eq:opt}, where $D(\cdot|\cdot)$ is a divergence function (here $L_2$ norm). Sparsity is controlled through the term $\mu \lVert\mathbf{H}\rVert_1$.
\begin{equation}\label{eq:opt}
    \bar{\mathbf{W}},\bar{\mathbf{H}} = \argmin_{\mathbf{W},\mathbf{H}} D\left(\mathbf{X}|\mathbf{WH}\right) + \mu \lVert\mathbf{H}\rVert_1
\end{equation}


%

\section{Segmentation evaluation}


\subsection{Experimental protocol}

Our model is trained on the full training set listed in Table~\ref{tab:datasets}. 
We use the train/dev/test partitions described in the associated articles.
When an annotation is not available for a given classification task, \textit{e.g.} music with DiHard data, the BCE loss is not considered for this specific class.
Models are randomly initialized and trained with the ADAM optimizer with an initial learning rate set to $10^{-3}$, no scheduler, and default other parameters.
The batch size is set to 64.
The optimization objective in eq.~\ref{eq:opt_obj} relies on a combination of $(\alpha, \beta, \gamma)$, but we decided to investigate only the impact of the spectrogram reconstruction via NMF loss weight $\beta$.
The model is trained for 36 hours on a single RTX8000 GPU card.
The subset used to pre-train $\mathbf{W}$ is composed of 1200 audio segments containing each class of interest, randomly sampled from the Albayz\'in train set.

\subsection{Segmentation results}

The performances are reported in Table~\ref{tab:seg_pef} for each binary segmentation task in terms of F1-score (and confidence intervals) on AragonRadio and DiHard test subsets.
The results are compared to the black-box model from \cite{lebourdais2024multilabel} and the NMF student from \cite{mariotte2024explainable} retrained on our data.
As shown in the previous study \cite{mariotte2024explainable}, the proxy model is on par with or outperforms the original black-box model on each task.

The last rows of the table present the results obtained with our models and different $\beta$ weights on the NMF loss.
In the $\beta=5$ scenario, the model offers poor OSD performance (F1=49.6\%) and the ND and MD performances remain slightly lower than the original teacher system. 
Only a slight improvement in SAD can be observed on the AragonRadio dataset (97.2\%).
By reducing the $\beta$ weight, the segmentation performance increases.
$\beta=0,1$ scenarios offer on-par performance on each segmentation task except OSD.
In the first case ($\beta=1)$, OSD performance remains lower than both baselines with 57.0\%.
In the second case ($\beta=0)$, the model reaches significantly similar OSD performance than student NMF (61.1\%).

Increasing the influence of spectrogram reconstruction during the training process tends to degrade the final segmentation performance.
This degradation could be explained by the additional constraint added to the system during training.
By increasing the reconstruction term, it makes the optimization of the reconstruction more difficult for the system.
It seems difficult for the system to find an optimal operating point concerning BCE and reconstruction terms.
By fully relaxing the reconstruction term, we obtain the best OSD score, but this probably limits the interpretation of the hidden representation $\mathbf{H}$.

\begin{figure*}[h!]
    \centering
    \includegraphics[trim=4cm 0.6cm 4cm 0.8cm, clip, width=0.95\textwidth]{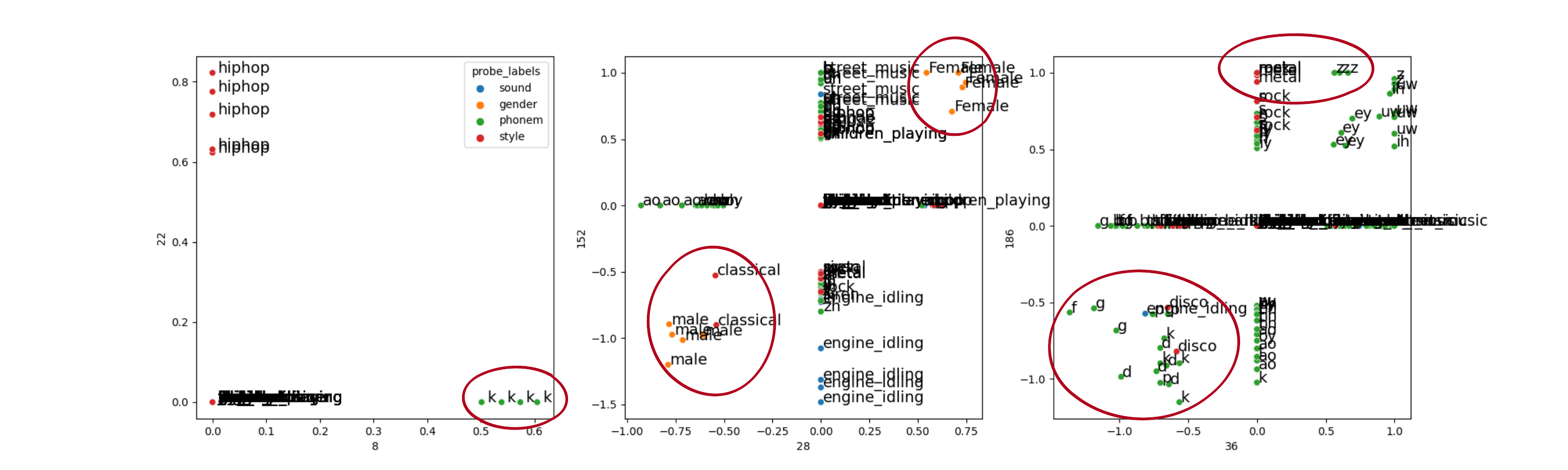}
    \caption{Visualization of some components with respect to audio classes: disentangled (left) or complementary (middle, right)}
    \label{fig:2D-plots}
    \vspace{-10pt}
\end{figure*}



\section{Probing H activations}
This section explores the $\mathbf{H}$ matrix and assesses whether it contains structured, fine-grained, and generic information.
The model used for this section has $\beta = 5$.
\vspace{-5pt}
\subsection{Experimental protocol}

The $\mathbf{H}$ matrix is analyzed by probing it on four classification tasks.
The performances are not expected to be close to the current state-of-the-art, but to inform us on how much the frozen $\mathbf{H}$ is relevant for a given task. 
We propose to probe $\mathbf{H}$ using linear layers similar to the $\Theta$ function. 
These probes require 1h of training on a single RTX6000 GPU card.
Our method needs a constant length inside a probe, the batch samples are thus zero-padded for training. The four tasks are described below.

\begin{table}[t]

    \centering
    \caption{Classification results (Accuracy and Unweighted Average Recall) for the 4 probes on the evaluation subset. 
    }
    \begin{tabular}{lcccc}
        \toprule
        Probing & Nb class & Acc (\%) & UAR & Rand (\%)\\ \midrule
        Phoneme & 39 & 50.6 & 30.2 & 2.6\\
        Music genre & 10& 55.1 &55.1 &10\\
        Speaker gender & 2 & 92.4 & 89.6 &50\\
        Sound events & 10 & 61.4 &60.7 &10\\\bottomrule
    \end{tabular}
    \label{tab:classif_probes_res}
    \vspace{-7pt}
\end{table}

\noindent\textbf{Music genre:} The GTZAN dataset is a balanced corpus designed for music genre classification with 100 excerpts of 30s for 10 classes, blues, classical, country, disco, hip-hop, jazz, metal, pop, reggae, and rock. 
We are aware of the limitations of this corpus, expressed in~\cite{sturmGTZANDatasetIts2014a} but as our objective is not performance, we decided to use it to show if the information available in $\mathbf{H}$ is enough to discriminate music genre.  

\noindent\textbf{Phoneme recognition:} TIMIT dataset~\cite{Timit_1993} is a standard of phonetic transcription tasks in English and contains sentences from North American speakers phonetically aligned. The set of phonemes used contains 61 phonemes with 20 vowels, 7 semi-vowels, 25 consonants, and 9 other symbols, mainly silence. We reduce the number of classes to 39~\cite{ohHierarchicalPhonemeClassification2021}.

\noindent\textbf{Gender classification (binary):} 
We use the French speakers from the Common Voice corpus from version 11.0 with the intended partition in train/validation and test. Common Voice contains sentences read by voluntary speakers from different backgrounds.

\noindent\textbf{Sound event classification:} The UrbanSound8k dataset~\cite{salamonDatasetTaxonomyUrban2014} contains 18.5h of audio extracted from freesound.org distributed into 10 audio event classes: \textit{air conditioner}, \textit{car horn}, \textit{children playing}, \textit{dog bark}, \textit{drilling}, \textit{engine idling}, \textit{gun shot}, \textit{jackhammer}, \textit{siren}, and \textit{street music}.
\vspace{-5pt}

\subsection{Classification results}
\vspace{-5pt}
Table~\ref{tab:classif_probes_res} contains the classification results obtained for each probing task. 
All of our probes largely outperform random results, even if they don't reach state of the art performances (almost 80\%).
We conclude that $\mathbf{H}$ includes fine-grained representations and not only segmentation-specific information, even if the representation has not been trained for these tasks.
The $\mathbf{H}$ representation not only leads to high-quality audio segmentation but also leads to acceptable probing results. 
%
As discussed in section \ref{sec:related}, a ``good'' representation must satisfy some specific properties.
This section demonstrates that probing the matrix $\mathbf{H}$ on different tasks achieves classification performances clearly above chance. This ensures that the matrix is \textbf{informative}: the factors (here classes) can be predicted from the components.

\section{Analysis of relevant components}
In this section, we deeply investigate how this matrix is structured with respect to interpretable factors, \textit{i.e.} probing classes.

\subsection{Relevant component extraction for explanation}\label{sec:relevant}

The segmentation prediction is obtained from $\mathbf{H}$ embedding with the $\Theta$ linear transformation.
To identify the most relevant NMF components, we first apply a pooling operation by averaging it over the time dimension: $z_k = \frac{1}{T}\sum_{t=1}^T h_{k,t}$.
Then, we define a filtered relevance vector $\mathbf{r}_{k,i} = z_k \times \theta_{k,i}$ if $r_{k,i} > \tau$, 0 otherwise, where $\theta_{k, i}$ is the $k$-th weight of the linear layer associated to audio sample $i$.

We have randomly selected 5 audio samples of each class used in the 4 probes, \textit{i.e.} a total of 60 classes (interpretable factors).
For each sample, we extract the relevant components $r_{ki}$.
The relevance vector is then normalized between 0 and 1 and a threshold $\tau = 0.5$ is applied to get binary values $b_{ki}$.
$b_{ki}=1$ means that the component $k$ is active for audio sample $i$.
18 components (7\%) are inactive whatever the audio samples.



\subsection{Compactness and modularity}


We want to check if $\mathbf{H}$ is \textbf{disentangled}, \textit{i.e.} to estimate how much each component $k$ explains a single factor $f$ (\textbf{modularity}), considering in this preliminary experiment, only 1D subspaces.
Then, we count the number of samples $n_k = \sum_i b_{ki}$ for which the component $k$ is active.
If the components are disentangled, $n_k$ should be around 5, as we have 5 samples per audio class.
We found that the number of components for which $4 \leq n_k \leq 6$ is 23 ($\simeq 9$\% modular components). 
Among them, component 8 is active for \phoneme{k} only, while component 22 is active for hip-hop music only as shown in Fig.~\ref{fig:2D-plots} (left).
All other components activate at least more than one class.
It means that the information embedded in one component can be either class-specific (modularity) or spread among the different classes.
We remind here, that the matrix $\mathbf{H}$ has not been trained, nor adapted, to the classes.

Finally, \textbf{compactness} is a good characteristic for interpretable dimensions: a factor is represented by a few components.
To verify this aspect, we count the number of components $m_i = \sum_k b_{ki}$ which are active for sample $i$.
We find that 57.3\% of the audio samples are represented by $m_i \leq20$ active components (7.8\%).
For example, gender information is encoded by at least components 28 and 152, but these two components do not embed the same information as shown in Fig.~\ref{fig:2D-plots} (middle). 
Component 28 also positively encodes \textit{pop music}, and negatively encodes \phoneme{O}.
Component 152 also positively encodes \phoneme{U,N,D,p,b,g} and \textit{hip hop}, and negatively encodes \textit{engine idling}, \phoneme{dZ, v} and \textit{metal}.
Higher level structured information seems to be embedded in $\mathbf{H}$.
Fig.~\ref{fig:2D-plots} (right) shows that plosives are positively associated with \textit{disco} which means that components 36 and 186 potentially encode drum rhythmic sounds.
Also, we can see that \phoneme{z} is associated with \textit{metal}. We hypothesize that component 186 might encode high-frequency sounds typical of guitar distortion.





This preliminary analysis leads us to think that the matrix $\mathbf{H}$ is hierarchically structured with different explanatory factors.
Some components are disentangled according to audio classes, but others are shared among them.
The audio classes derived from the probing task are probably not fine-grained enough to be considered as atomic interpretable factors.
We also checked that the matrix meets the compactness property.
Of course, a deeper investigation is needed to interpret all components.

\section{Conclusion and perspectives}
In this paper, we have investigated a new explainable by-design audio segmentation model based on NMF, which provides simultaneously multiple labels at the frame level.
We demonstrated that this model reaches similar segmentation performances on standard datasets in comparison to state-of-the-art models.
%
We have also shown that the embedding matrix $\mathbf{H}$ is a good candidate for the extraction of interpretable representation.
Through the use of probing, we confirmed that this matrix not only contains information specific regarding the four target classes it has been trained for but also embeds fine-grained informative components required to discriminate phonemes, sound events, gender, or music genres.
%
Moreover, a preliminary analysis of this matrix highlights a form of disentanglement by demonstrating that 9\% components encode a unique probe class (modularity). 
We also found that 57.3\% of the samples were represented by less than 7.8\% of the components.
To conclude, the matrix is hierarchically structured with explanatory factors and meets informativeness, modularity, and compactness.

There is a need for more atomic, fine-grained explanatory factors than the proposed probe classes
, and also higher dimensional interpretable subspaces.
Another perspective would be the exploration of the temporal information that is currently lost in the matrix analysis but is crucial for segmentation.



\section{Ackowledgments}
This project has received funding from the European Union’s Horizon 2020 research and innovation program under the Marie Skłodowska-Curie grant agreement No 101007666 (Esperanto project). 
This work was performed using HPC resources from GENCI–IDRIS (Grant 2022-AD011012565).

\bibliographystyle{IEEEbib}
\bibliography{Odyssey2024_BibEntries}

\end{document}